\documentclass[pre,aps,amssymb,showpacs,superscriptaddress]{revtex4}
\usepackage[dvips]{epsfig}

\begin{document}

\title{The phase diagram of the multi-dimensional Anderson localization via
analytic determination of Lyapunov exponents}

\author{V.N.~Kuzovkov}

\affiliation{Institute of Solid State Physics, University of
Latvia, 8 Kengaraga Street, LV -- 1063 RIGA, Latvia}

\email{kuzovkov@latnet.lv}

\author{W. von Niessen}
\affiliation{Institut f\"ur Physikalische und Theoretische Chemie,
Technische Universit\"at Braunschweig, Hans-Sommer-Stra{\ss}e 10,
38106 Braunschweig, Germany}

\date{Received \today}

\begin{abstract}
The method proposed by the present authors to deal analytically
with the problem of Anderson localization via disorder [J.Phys.:
Condens. Matter {\bf 14} (2002) 13777] is generalized for higher
spatial dimensions D. In this way the generalized Lyapunov
exponents for diagonal correlators of the wave function, $\langle
\psi^2_{n,\mathbf{m}} \rangle$, can be calculated analytically and
exactly. This permits to determine the phase diagram of the
system. For all dimensions $D > 2$ one finds intervals in the
energy and the disorder where extended and localized states
coexist: the metal-insulator transition should thus be interpreted
as a first-order transition. The qualitative differences permit to
group the systems into two classes: low-dimensional systems
($2\leq D \leq 3$), where localized states are always
exponentially localized and high-dimensional systems ($D\geq
D_c=4$), where states with non-exponential localization are also
formed. The value of the upper critical dimension is found to be
$D_0=6$ for the Anderson localization problem; this value is also
characteristic of a related problem - percolation.
\end{abstract}


\maketitle

\section{Introduction}
\subsection{Experiment and theory}

Disorder leads to important physical effects which are of quantum
mechanical origin. This has been revealed by Anderson
\cite{Anderson} in the study of a disordered tight-binding model. This problem
has attracted great attention over many decades. A breakthrough
came with the scaling theory of localization \cite{Abrahams}. All
states in a one-dimensional system (1-D) are localized, whereas in
3-D a metal-insulator transition occurs. An analytic solution is
only known for the 1-D problem \cite{Molinari}. Although there was
no general analytical solution available, there was consensus that
in 2-D all states are localized.

For quite some time after the advent of the scaling theory, many
believed it to be essentially under control. This view is less
secure nowadays, in part because recent experiments have
challenged conventional wisdom about disordered 2-D systems. The
2-D case still presents a problem which has become apparent by
experiments \cite{Abrahams2,Kravchenko,Ilani1,Ilani2}. These
experiments are still being discussed controversially.
Experimental reality is certainly more complex than a simple
tight-binding model, but these results provide a good reason for
revisiting this classic theoretical problem.

Recently we have been able to solve the 2-D case analytically
\cite{Kuzovkov}. We have shown that in principle there is the
possibility that the phase of delocalized states exists for a
\textit{non-interacting electron system}. For energies and
disorder, where extended states may exist we find a coexistence of
these localized and extended states. Thus the Anderson
metal-insulator transition exists and should be regarded as a
first order phase transition. Consequently we have returned to the
old idea of Mott \cite{Mott73,Lee85} that the metal-insulator
transition is discontinuous. This alternative idea was in its
history completely abandoned with the advent of the scaling theory
of localization. Thus we expect a renaissance of it.

This result implies the failure of the scaling theory of Anderson
localization. Although this paper is published \cite{Kuzovkov} and
constitutes the basis for a new analytical investigation of the
Anderson problem in the present paper for the case of higher
dimensions (N-D problem) one has to accept the following: (i) the
paper \cite{Kuzovkov} requires an independent confirmation, which
requires a certain time; (ii) in the history of the problem one
has developed quite a few conceptions and this leads to a critical
attitude towards the new theory - i.e. there is a resistance - as
it departs from conventional wisdom. This asks for a critical
evaluation of the new theory and its results \cite{Comment,Reply}.
One should acknowledge that the problem is many-sided and quite
complex. Thus logical errors are possible which do not lie at the
surface. Perhaps a reference to B. Pascal is appropriate here:
\textit{A truth is so delicate that any small deviation from it
leads you to a mistake, but this mistake is also so delicate that
after a small retreat you find yourselves in a truth again.}
Without an exact analytic solution a discussion cannot lead to
firm results.

Let us add at this point some reflections of a historical nature.
It is well known that exact results in the field of phase
transitions are exceedingly rare (so to say they are exceptions)
and they exist only under certain conditions and limitations. The
situation in the field of Anderson localization shows in this
respect a certain similarity to the history of the theory of the
second order phase transitions and of the Ising model. The
phenomenological theoretical work by Landau \cite{Landau} and the
numerical work by Bethe \cite{Bethe} were prior to Onsager's
analytical theory \cite{Onsager} and they were well accepted. But
the results of the analytical theory departed strongly from the
phenomenological and numerical results in that it gave different
critical exponents. But there are limitations to the analytical
theory. The exact solution of Onsager for the Ising model is only
possible for 2-D, without magnetic field and only for the case of
an interaction between nearest neighbours. This is also the case
in our present investigation. An exact solution is only possible
(a) for the tight-binding approximation with (b) diagonal
disorder, where (c) on-site potentials  are independently and
identically distributed, i.e. for the conventional Anderson model.

A contradiction between results from phenomenological theories
(Landau, scaling theory) and analytical results (Onsager, the
present work) have (Onsager) or might (the present case) lead to
scientific advances. The metal-insulator transition is a phase
transition. In the case of such phase-transitions the results tend
to be unstable and depend on the approximations. An exact solution
thus rarely shows an agreement with results from a
phenomenological theory and numerical work, because the two latter
methods make use of additional approximations and assumptions,
whose consequences are hardly controllable. To return to the
advances made possible by Onsager's analytical theory: V. Ginzburg
got in the year 2003 the nobel prize for the work (together with
L. Landau \cite{Ginzburg}), in which he used the phenomenological
theory of second order phase transitions. This all means simply
that phenomenological and numerical results frequently require
certain corrections, which is only then possible, when an exact
solution is known.

The main aim of the present article is thus the generalization of
the mathematical tools of the previous article\cite{Kuzovkov} to
the case of higher dimensional spaces and a physical
interpretation of the new results.

\subsection{Structure of the present article}
The outline of the present article is a follows.
The article \cite{Kuzovkov} opens the possibility to determine
the phase diagram of the multi-dimensional Anderson localization
analytically. This is based on the analytic calculation of Lyapunov
exponents. The argumentation is given at the beginning of
chapter 2 in detail. There we also give a short derivation
of the equations for the N-D Anderson localization
problem, and a necessary summary of the results of the article
\cite{Kuzovkov}. The connection between the Anderson localization
problem and signal theory is discussed and the most important
concept of the proposed method - the filter $H(z)$ - is defined.
To understand fully these aspects of the theory  a knowledge of
the first paper \cite{Kuzovkov} is recommended. The filter $H(z)$
is generalized for higher dimensions D of the space. The
investigation of its properties and the corresponding physical
interpretation constitute the content of several chapters. The
theory for high-dimensional systems ($D\geq 4$) is presented in
chapter 3, whereas the theory for low-dimensional systems is given
in chapter 4. An appendix deals with the mathematical conditions
for the physical interpretation of the filter $H(z)$ and
a more detailed discussion of related aspects of the problem.
Because of the length of the present article we omit both the
mathematical details of the theory as well as an extended
discussion of the implications of the results. This will
be given in a future publication, where also a different
procedure for part of the derivation is used instead of an
average over initial conditions as done here.

\section{New methods}

\subsection{Anderson localization and generalized diffusion}

Recently we have been able to solve the 2-D case analytically
\cite{Kuzovkov}. The tight-binding equation in 2-D is solved for
the wave function $\psi _{n,m}$ and the second moments (diagonal
correlators $\langle \psi^2_{n,m} \rangle$):
\begin{equation} \label{recursion D}
\psi _{n+1,m}=(E-\varepsilon _{n,m}) \psi _{n,m}-\psi_{n-1,m} -
\sum_{m^{\prime}}\psi _{n,m+m^{\prime}},
\end{equation}
where the summation over $m^{\prime}$ runs over the nearest
neighbours of site ($n,m$) in layer $n$ that is in a space of
dimension $p=D-1=1$. We assume taking the limit to infinite size
$L \rightarrow \infty$ in $p$-space. The equation is solved with
an initial condition
\begin{equation}\label{rand D}
\psi _{0,m}=0,  \psi _{1,m}=\alpha_{m}.
\end{equation}
The on-site potentials $ \varepsilon_{n,m} $ are independently and
identically distributed with existing first two moments,
$\left\langle \varepsilon _{n,m}\right\rangle =0$ and
$\left\langle \varepsilon _{n,m}^2\right\rangle =\sigma ^2$.


The mathematical idea shortly discussed above is based on a simple
physical idea. The aim of the article \cite{Kuzovkov} consisted in
determining a \textit{phase diagram} for the Anderson localization
problem. To this purpose the concept of the Lyapunov exponent
$\gamma$ \cite{Kuzovkov} is extremely well suited. The Lyapunov
exponent plays the role of an order parameter: $\gamma\equiv 0$
for the metallic phase and $\gamma \neq 0$ for the insulating
phase. The value $\gamma \neq 0$ for the insulating phase implies
a divergence of certain mean values as a function of the index
$n$. If we understand what this divergence means it will become
clear, which quantities have to be calculated.

This divergence is a general property of a number of different
stochastic systems. To give a clear picture we consider in the
following the equations for 1-D systems and in particular for
diffusion (random walks)
\begin{eqnarray}\label{recursion2}
\psi_{n+1}=\psi_n +\varepsilon_n ,
\end{eqnarray}
and Anderson localization
\begin{eqnarray}\label{recursion}
\psi_{n+1}=(E-\varepsilon_n)\psi_n-\psi_{n-1}.
\end{eqnarray}
For symmetric diffusion, $\left\langle \varepsilon_n
\right\rangle = 0$, $\left\langle \varepsilon^2_n \right\rangle =
\sigma^2$, the simplest and most important characteristic
is the second moment  $\left\langle \psi^2_n \right\rangle = \psi^2_0+\sigma^2
n$. A divergence of this variable for $n \rightarrow \infty$
is an unambiguous proof for the existence of diffusion. It is also
analytically possible to calculate other even moments, e.g. $\left\langle
\psi^4_n \right\rangle$. These, however, do not supply new information.
All even moments diverge simultaneously, and
they do this as a power of $n$. A similar divergence also occurs for other
mean values of the functions of  $|\psi_n|$, e.g.
$\left\langle |\psi_n| \right\rangle$; however, these can only be calculated
numerically. In other words, for the determination of the existence of
diffusion it is fully sufficient to calculate only one moment, the second moment.

The equation (\ref{recursion}) is nothing else but a
generalization of equation (\ref{recursion2}). The mean value
$\left\langle \psi^2_n \right\rangle = f(n)$ describes in this
case generalized diffusion. If $f(n)$ is bounded,  $f(n) < \infty
$, then the proper dynamics in equation (\ref{recursion}) is
stable and we have no diffusion (Lyapunov exponent $\gamma \equiv
0$). This corresponds in a physical interpretation to the
existence of delocalized states. A divergence of the function
$f(n)$ for $n \rightarrow \infty$ corresponds to generalized
diffusion (localized states); in this case one could distinguish
in addition between non-exponential localization ($f(n)$ is a
non-exponential function) and exponential localization ($f(n)
\propto \exp(2\gamma n)$ with $\gamma \neq 0$).

It is quite important to stress here that the formation of diffusion in
a dynamic system represents always a type of phase transition.
Therefore one can easily find the corresponding critical values and determine
the phase diagram of the system. As for proper diffusion, an analytic
calculation for eq. (\ref{recursion})
is only possible for even moments \cite{Molinari,Kuzovkov}.

For these reasons it is fully sufficient to consider only the
equations for the second moments (correlators) to determine the
phase diagram of the system \cite{Kuzovkov}. This idea has a
certain similarity to the method of Pendry \cite{Pendry82}. Pendry
has introduced the density matrix to the problem of waves in
disordered systems. However, he was of the opinion (without giving
a proof), that for Anderson localization all higher-order density
matrices and hence all higher moments need to be considered. This
program is practically unfeasible, so Pendry limited himself in
\cite{Pendry82} to the second moment. Instead of solving the
equations for the second moments exactly Pendry has mainly studied
the properties of the transfer matrix and the process of its
diagonalization. For the transfer matrix, however, the
investigation of the thermodynamic limit is impossible (see
discussion in \cite{Reply}), and consequently in this way the
phase diagram of the system cannot be obtained.

\subsection{Equations for correlators}

In the present paper we present the analytic solution for the
general case $D>2$. The knowledge of paper \cite{Kuzovkov} is
prerequisite for understanding the present one as it contains the
full formalism. The generalization of this formalism to the N-D
case presents no problem (see below). The main
eqs.~(\ref{recursion D}),(\ref{rand D}) remain valid, only scalar
quantities become vector quantities.

The tight-binding equation in $1+p$ dimension is (primed indices
are summed)
\begin{equation} \label{recursion DN}
\psi _{n+1,\mathbf{m}}=-\varepsilon _{n,\mathbf{m}} \psi
_{n,\mathbf{m}}-\psi_{n-1,\mathbf{m}} +
\mathcal{L}_{\mathbf{m},\mathbf{m^{\prime}}}\psi
_{n,\mathbf{m^{\prime}}} ,
\end{equation}
\begin{eqnarray}\label{L}
\mathcal{L}_{\mathbf{m},\mathbf{m^{\prime}}}=E\delta_{\mathbf{m},\mathbf{m^{\prime}}}
-\sum_{\mathbf{m^{\prime \prime}}}\delta_{\mathbf{m+m^{\prime
\prime}},\mathbf{m^{\prime}}} ,
\end{eqnarray}
(summation over $\mathbf{m^{\prime \prime}}$ runs over the nearest
neighbours) with initial condition $\psi_{0,\mathbf{m}}=0$ and
$\psi_{1,\mathbf{m}}=\alpha_{\mathbf{m}}$.

One introduces the correlators (the averages are taken over
disorder)
\begin{eqnarray}
x(n)_{\mathbf{m},\mathbf{l}}=\left\langle \psi _{n,\mathbf{m}}\psi
_{n,\mathbf{l}} \right\rangle , \\
y(n)_{\mathbf{m},\mathbf{l}}=\left\langle \psi _{n,\mathbf{m}}\psi
_{n-1,\mathbf{l}} \right\rangle .
\end{eqnarray}

From the eq.(\ref{recursion DN}) one easily obtains the system of
equations:
\begin{eqnarray}
x(n+1)_{\mathbf{m},\mathbf{l}}=\delta_{\mathbf{m},\mathbf{l}}\sigma^2
x(n)_{\mathbf{m},\mathbf{l}}+x(n-1)_{\mathbf{m},\mathbf{l}}+\\
\nonumber
\mathcal{L}_{\mathbf{m},\mathbf{m^{\prime}}}x(n)_{\mathbf{m^{\prime}},\mathbf{l^{\prime}}}
\mathcal{L}_{\mathbf{l^{\prime}},\mathbf{l}}  -
\mathcal{L}_{\mathbf{m},\mathbf{m^{\prime}}}y(n)_{\mathbf{m^{\prime}},\mathbf{l}}-
\mathcal{L}_{\mathbf{l},\mathbf{l^{\prime}}}y(n)_{\mathbf{l^{\prime}},\mathbf{m}}, \\
y(n+1)_{\mathbf{m},\mathbf{l}}=-y(n)_{\mathbf{l},\mathbf{m}}+
\mathcal{L}_{\mathbf{m},\mathbf{m^{\prime}}}x(n)_{\mathbf{m^{\prime}},\mathbf{l}}
.
\end{eqnarray}
They can be solved explicitly by introducing of the
Z-transform\cite{Weiss}
\begin{eqnarray}
X(z)_{\mathbf{m},\mathbf{l}}=\sum_{n=0}^{\infty}\frac{x(n)_{\mathbf{m},\mathbf{l}}}{z^n}
, \\
Y(z)_{\mathbf{m},\mathbf{l}}=\sum_{n=0}^{\infty}\frac{y(n)_{\mathbf{m},\mathbf{l}}}{z^n}
,
\end{eqnarray}
which turn the equations into
\begin{eqnarray}
(z-z^{-1}-\sigma^2\delta_{\mathbf{m},\mathbf{l}})X(z)_{\mathbf{m},\mathbf{l}}
-x(1)_{\mathbf{m},\mathbf{l}}=
\mathcal{L}_{\mathbf{m},\mathbf{m^{\prime}}}X(z)_{\mathbf{m^{\prime}},\mathbf{l^{\prime}}}
\mathcal{L}_{\mathbf{l^{\prime}},\mathbf{l}} \\ \nonumber -
\mathcal{L}_{\mathbf{m},\mathbf{m^{\prime}}}Y(z)_{\mathbf{m^{\prime}},\mathbf{l}}-
\mathcal{L}_{\mathbf{l},\mathbf{l^{\prime}}}Y(z)_{\mathbf{l^{\prime}},\mathbf{m}} ,\\
zY(z)_{\mathbf{m},\mathbf{l}}=-Y(z)_{\mathbf{l},\mathbf{m}}+
\mathcal{L}_{\mathbf{m},\mathbf{m^{\prime}}}X(z)_{\mathbf{m^{\prime}},\mathbf{l}}
.
\end{eqnarray}
Iteration of the second equation yields
\begin{eqnarray}
Y(z)_{\mathbf{m},\mathbf{l}}=\frac{z}{z^2-1}
\mathcal{L}_{\mathbf{m},\mathbf{m^{\prime}}}X(z)_{\mathbf{m^{\prime}},\mathbf{l}}
-\frac{1}{z^2-1}X(z)_{\mathbf{m},\mathbf{l^{\prime}}}\mathcal{L}_{\mathbf{l^{\prime}},\mathbf{l}}
.
\end{eqnarray}
We then obtain an equation for $X(z)_{\mathbf{m},\mathbf{l}}$
only, which can be solved by double Fourier expansion.
A complete presentation of the method is rather lengthy and will
thus be the subject of a following paper. An shortcut which quickly leads
to the correct equations is suggested in section 3.2 of \cite{Kuzovkov}):
averaging over translations in $p$-space of boundary conditions,
$\alpha_{\mathbf{m}}$.

Upon averaging over an ensemble of initial conditions (as
described in \cite{Kuzovkov}) such that
$\overline{\alpha_{\mathbf{m}}\alpha_{\mathbf{m^{\prime}}}}=\Gamma_{{\mathbf{m-m^{\prime}}}}$,
the problem is translation-invariant in transverse directions. We
then put:
\begin{eqnarray}
X(z)_{\mathbf{m},\mathbf{l}}=\int\frac{d^p\mathbf{k}}{(2\pi)^p}
X(z,\mathbf{k})e^{i\mathbf{k}(\mathbf{m}-\mathbf{l})} .
\end{eqnarray}
After averaging over initial conditions the diagonal correlator
becomes independent of $\mathbf{m}$:
\begin{eqnarray}
x(n)_{\mathbf{m},\mathbf{m}} \equiv x_n , \\
X(z)_{\mathbf{m},\mathbf{m}} \equiv X(z)=
\int\frac{d^p\mathbf{k}}{(2\pi)^p} X(z,\mathbf{k}).
\end{eqnarray}
We obtain the final equations
\begin{eqnarray}
\frac{(z-1)}{(z+1)}[w^2-\mathcal{E}^2(\mathbf{k})]X(z,\mathbf{k})=
\Gamma (\mathbf{k})+\sigma^2X(z) ,\\
\mathcal{E}(\mathbf{k})=E - 2\sum_{j=1}^p\cos (k_j), \label{Ek}  \\
w ^2=\frac{(z+1)^2}z ,\label{gamma1}
\end{eqnarray}
or
\begin{eqnarray}
 X(z)=H(z)X^{(0)}(z) ,\label{X} \\
X^{(0)}(z)=\frac{(z+1)}{(z-1)}\int\frac{d^p\mathbf{k}}{(2\pi)^p}
\frac{\Gamma (\mathbf{k})}{[w^2-\mathcal{E}^2(\mathbf{k})]} ,\\
\frac{1}{H(z)} = 1-\sigma^2
\frac{(z+1)}{(z-1)}\int\frac{d^p\mathbf{k}}{(2\pi)^p}
\frac{1}{[w^2-\mathcal{E}^2(\mathbf{k})]} ,\label{H}
\end{eqnarray}
where the $X^{(0)}(z)$ (or $ x_n^{(0)}$) refer to the ordered
system ($\sigma \equiv 0$) and the $X(z)$ (or $x_n$) to the
disordered one ($\sigma \neq 0$). For eq.(\ref{X}) the inverse
Z-transform gives the convolution property \cite{Weiss}:
\begin{equation}\label{xn0}
x_n=\sum_{l=0}^n x_l^{(0)} h_{n-l},
\end{equation}
with
\begin{eqnarray}
  H(z)=\sum_{n=0}^{\infty}\frac{h_n}{z^n} .
\end{eqnarray}
\subsection{Anderson localization and signal theory}

The essential point in the analysis with respect to the localized
or extended character of the states is to make use of signal
theory \cite{Weiss} from electrical engineering and switch from an
investigation of the moments $x_n$ to an analysis of the filter
functions $h_n$.

In the theory of signals \cite{Weiss}, $x_n^{(0)}$ (or
$X^{(0)}(z)$) is the input signal, and $x_n$ (or $X(z)$) is the
output. Asymptotic behaviour of the solution is completely
determined by the filter $h_n$ (or $H(z)$). The concept of the
system function is a general and abstract description of the
problem of localization. Thus the filter function has to be
analysed to obtain general results and not the multitude of
signals. This has been done in \cite{Kuzovkov} for the 2-D case.
This procedure has a certain similarity with the transition to an
operator formalism in quantum mechanics. Particular signals $x_n$
depend on the initial conditions used and do not carry much
physical information because of the unconventional normalization
\cite{Kuzovkov}. For the localization problem the only property
that matters is whether a signal belongs to the bounded or
unbounded class and this can be derived from the filter.

The essence of localization is contained in the filter $H(z)$. We
study the filter $H(z)$ with properties described by generalized
Lyapunov exponents \cite{Kuzovkov,Pendry}. The filter is a
fundamental function of the disorder $\sigma$ only.

A filter $h_n$ is uniquely characterized by a pole-zero diagram of
its image $H(z)$ which is a plot of the locations of the poles
$\lambda_i$ and zeros of $H(z)$ in the complex-$z$ plane. We
provide just a brief summary here, for more details consult
\cite{Kuzovkov,Weiss}. The signals $x_n^{(0)}$ and $x_n$ are real,
therefore $H(z)$ will have poles and zeros that are either on the
real axis, or come in conjugate pairs. For the inverse
$Z$-transform $H(z)\Rightarrow h_n$ one needs to know the
\textit{region of convergence} (ROC). Physical considerations
dictate that only \textit{causal} filters ($h_n=0$ for $n<0$)
should be considered. They have ROCs outside a circle that
intersects the pole with $\max{|\lambda_i}|$.  A causal filter is
\textit{stable} (bounded input yields a bounded output) if the
\textit{unit circle} $|z|=1$ is in the ROC. Note that the explicit
calculation of $h_n$ by the inverse $Z$-transform is not
necessary, and it is also not feasible analytically due to the
complexity of the function $H(z)$. Only the type of the filter --
stable or unstable -- needs to be determined. The delocalized
states (bounded output) are obtained by transforming the physical
solutions inside the band (bounded input) provided that the filter
$H(z)$ is stable. Seeking for poles is quite a simple analytical
task which gives rather general results by elementary methods.

As an example for the general and abstract description of the
problem of localization as stated above let us consider the
following problem. It is well-known that disorders broadens the band;
new states outside the old band arise for $|E|>E_b=2D$.
Are among these new states also extended states? Numerical work for $D=3$
\cite{Kramer,Grussbach95} ascertains this. This is the socalled
reentrant behaviour of the mobility edge: the change from
localized to extended states and back to localized ones upon
increasing the disorder occurs for certain fixed energies. Because
in the literature there is no physical explanation for this phenomenon
one has simply accepted these results without critically examining them.

With the help of signal theory we have found a particular
transformation, which gives a connection between the states in an
ideal system (zero disorder) and states in a system with disorder.
Extended states (zero disorder) as input (bounded) signal
transform into localized states (nonzero disorder) as output
(unbounded) signal if the filter which is responsible for this
transformation is unstable. If the filter is stable, then extended
states (bounded input signal) transform into extended states
(bounded output signal). It is known that for zero disorder
outside the band, $|E| > E_b$, there do exist only mathematical
solutions which cannot be normalized. These correspond to an
unbounded input signal. It is impossible to find a filter which
permits a transformation of the unbounded input signal
(mathematical solution) into a bounded output signal (extended
states in $|E| > E_b$). The reverse - the transformation of an
unbounded input signal (mathematical solution) into a unbounded
output signal (localized states in $|E| > E_b$) - is on the other
hand possible. I.e. the mathematical procedure developed by us
generates in this case a negative answer to the posed question.

We have shown in \cite{Kuzovkov} that the filter $H(z)$ is a
non-analytic function of the complex variable $z$; this result
remains valid also in the multi-dimensional case. The unit circle
$|z|=1$ divides the complex plane into two analytic domains: the
interior and exterior of the unit circle. The inverse Z-transform
is quite generally defined via contour integrals in the complex
plane
\begin{equation}\label{inverse}
h_n=\frac{1}{2\pi i}\oint H(z)z^n \frac{dz}{z} .
\end{equation}
and this definition is only possible in an analytic domain. In
this way  in the formal analysis of the problem multiple solutions
result. The first solution $H_{+}(z)$ is defined outside the unit
circle and always exists. The filter $H_{+}(z)$ describes
localized states and it is possible to connect its properties with
the notion of the localization length \cite{Kuzovkov}. The second
solution $H_{-}(z)$ is defined inside the unit circle and does not
always represent a solution which can be physically interpreted
(this is the mathematical consequence that the filter be causal).
The filter $H_{-}(z)$ describes delocalized states. The
coexistence of the two solutions was physically interpreted in
\cite{Kuzovkov,Reply} as the coexistence of two phases -- an
insulating and a metallic one. Then the metal-insulator transition
should be looked at from the basis of first-order phase transition
theory.

\subsection{Conformal mapping}

The $p$-dimensional integral on the r.h.s.\ of eq.~(\ref{H}) can
be reduced to a one-dimensional integral. Consider the identity
\begin{equation}\label{rechts}
\frac{ 1}{w ^2-\mathcal {E}^2(\mathbf{k})}=\int_{-\infty }^\infty
\frac{\delta \left( y+\sum_{j=1}^p2\cos (k_j)\right) dy}{w
^2-(E+y)^2}.
\end{equation}
The integral representation  of the Dirac $\delta$-function and the
Bessel function
\begin{equation}\label{J0}
  J_0(x)=\frac 1{2\pi }\int_{-\pi }^\pi e^{ix\cos (k)}dk.
\end{equation}
will be used. Following \cite{Kuzovkov}, let us define a complex
parameter $w =u + i v$ from eq. (\ref{gamma1}) in the upper
half-plane, $v  \geq 0$. Using the methods of complex variable
theory we get
\begin{equation}\label{rechts2}
\frac{ 1 }{{(2\pi )}^{p}}\int \frac{d \mathbf{k}}{w
^2-{\mathcal{E}}^2 (\mathbf{k})}= \frac 1{iw } Y_D(w ,E),
\end{equation}
where
\begin{equation}\label{Y}
  Y_D(w, E) = \int_0^\infty
\left[ J_0(2t)\right] ^{D-1} \cos(Et) \exp (iw t) \, dt.
\end{equation}
Changing the complex variable $z$ to the parameter $w$ corresponds
to the conformal mapping of the inner part [$|z| \leq 1$, $w =
-(z^{1/2}+z^{-1/2})$)] or the outer part [$|z| \geq 1$, $w =
(z^{1/2}+z^{-1/2})$] of the circle onto the upper half-plane. The
circle itself maps onto the interval $ [ -2,2 ]$. Note also that
if $H(z)=0$ has complex conjugate poles, then on the upper $w$
half-plane they differ only by the sign of $u$. To avoid
complicated notations, we seek for poles in the sector $u\geq 0, v
\geq 0$ and double their number if we find any.

The inverse function \cite{Kuzovkov}
\begin{equation}\label{z-inverse}
z=-1+\frac{w ^2}2\pm \frac w 2\sqrt{w ^2-4}
\end{equation}
is double-valued. Its branch with the minus sign maps the $w$
sector onto the inner part of the half-circle ($|z| \leq 1$). The
second branch with the plus sign gives a mapping onto the
half-plane with the half-circle excluded ($|z| \geq 1$). Therefore
in the parametric $w$-representation
\begin{equation}\label{zz}
    \frac{z+1}{z-1} =\pm \frac{w}{\sqrt{w^2-4}}
\end{equation}
and
\begin{equation}\label{Hpm}
  \frac{1}{H_{\pm}(z)}=1 \pm \sigma^2 i \frac{Y_D(w,E)}{\sqrt{w^2-4}} .
\end{equation}

\section{High-dimensional systems}

\subsection{Upper critical dimension}

It is generally assumed that 2-D systems mark the
borderline between high and low dimension \cite{Rice97}. The
existence of a transition in 3-D is not questioned
(high-dimensional systems). These assumptions, however, originate from
the scaling theory of localization. Here the marginal
dimension is $D_M = 2$, and a phase transition exists only
for $D > D_M$. Thus perturbation theory \cite{Abrahams2}
for $D=2+\varepsilon$ ($\varepsilon \ll 1$) is possible. The effect
of statistical fluctuations cause a change of regime at $D_c=4$
\cite{Lee85,Kunz83}; in this way the upper critical dimension for
localization $D_c$ arises. For $D > D_c$ there should not exist
a phase transition.

On the one hand these statements referring to higher dimensions
are numerically nearly impossible to ascertain \cite{Markos02}.
Statistics is bad and the length of the system $L$ very small.
Even for $D=3$ progress towards extracting reliable numerical
estimates of critical quantities has been remarkably difficult
\cite{Queiroz02}. On the other hand, if the results from the
scaling theory of localization for 2-D systems are faulty (this is
what we claim), then the corresponding division of the systems
into low and high dimensional ones is also wrong. Here one must
develop an alternative picture.

In the theory of critical phenomena \cite{Stanley,Ma} many
systems belong to a class, where an upper critical dimension
$D_0$ has a totally different physical meaning. It denotes the dimension,
from where on the mean field approximation is exact, or where with other words
all critical exponents reach stationary values.
I.e. in this case the phase transition does exist also for $D
> D_0$; only in the limit $D \rightarrow \infty$ the transition disappears,
simply because the corresponding critical values have gone to
infinity. Systems of quite different physical nature may have the
same value of the upper critical dimension $D_0$. E.g. one finds
$D_0 = 4$ not only for the theory of magnetism \cite{Stanley,Ma}
(in this case there exists also the marginal dimension, below
which no phase transition is possible), but also in kinetics
(cooperative phenomena in bimolecular processes by
diffusion-controlled reactions) \cite{Kuzovkov88,Kotomin96}; in
the latter case, however, there is no marginal dimension. Another
example is percolation, where the upper critical dimension is
$D_0=6$ \cite{Perc}.

\subsection{Stability and poles: solution $H_+(z)$}

Let us start first from a purely mathematical comment: the
integral $Y_D(w,E)$ is always finite for all $w=u+iv$ in the sector
$u \geq 0, v \geq 0$ for high-dimensional systems with $D\geq 4$.
This fact clearly follows from the asymptotic behaviour of the
Bessel function for large values of its argument, $J_0(2t)\approx
\frac{1}{\sqrt{\pi t}}\cos (2t-\pi /4)$. We shall further on see
that the dimensionality $D=D_c=4$ is critical for localization,
although this is no proper upper critical dimension $D=D_0$, which
we shall determine below. Let us consider therefore the properties
of the filter-functions in this region of the value of the
dimension of space.

Let us consider first the solution $H_+(z)$. According to
\cite{Weiss}, the ROC of the causal filter is defined by the
inequality $|z|>\max|\lambda_i|$, where $\lambda_i$ are the poles.
The case when the system function has a pole at $z=\lambda >1$ is
the simplest one for an interpretation. In terms of signal
theory \cite{Weiss} the filter $H_{+}(z)$  is unstable since the
pole lies outside the unit circle $|z|=1$ in the complex
$z$-plane. As shown in \cite{Kuzovkov}, such a filter describes
exponentially localized states (insulating phase). In order to see
this one can exploit a basic inverse Z-transform \cite{Weiss}:
\begin{equation}\label{basic}
H(z)=z/(z-\lambda )\Rightarrow h_n=\lambda ^n .
\end{equation}
Therefore the pole of $H_{+}(z)$ at $z=\lambda=\exp(2\gamma)$
leads to exponential growth of the system function $h_n$ which in
turn implies [eq.\ (\ref{xn0})] an exponentially increasing mean
squared amplitude $x_n$. The growth exponent $\gamma$ is the
so-called generalized Lyapunov exponent \cite{Molinari} related to
the localization length by $\xi =\gamma^{-1}$.

The value of the Lyapunov exponent $\gamma$  defines the phase.
We start from the mathematical definition that
all states with $\gamma \neq 0$ belong to the insulating phase.
The states with $\gamma \equiv 0$ on the other hand correspond to
a metal. According to this definition the states with
non-exponential localization  also belong to the metallic phase,
because they correspond to the value $\gamma \equiv 0$. We can
consider these states as a bad metal, in contrast to a good metal,
where one has truly delocalized states.

We would like to give here a summary of the results which emerge
from an analysis of the pole diagram (for details see the
Appendix). The function $Y_D(w,E)$ defined by eq.\ (\ref{Y}) is
purely imaginary for $v=0$, $u >u_0$,
\begin{equation}\label{u0}
u_0 = 2p+|E| ,
\end{equation}
and the system function $H_{+}(z)$ itself is real.   The pole must
be located (if present at all) exactly in this region of the
parameter $w$. It can be found as a solution of an in general
transcendental equation
\begin{equation}\label{om+}
  \sigma^2 \Omega _{+}(u)=1 ,
\end{equation}
 where
\begin{equation}\label{W+}
  \Omega _{+}(u) = \frac{1}{\sqrt{ u^2-4}} \int_0^\infty
\left[ J_0(2t)\right] ^{D-1} \cos(Et) \sin (u t)dt.
\end{equation}
For high-dimensional systems with $D\geq 4$ the function $\Omega
_{+}(u)$ has a maximum at $u =u_0$ and decreases monotonically for
$u > u_0$. The system function has a pole if the disorder exceeds
a critical value, $\sigma > \sigma_0(E)$, where
\begin{equation}\label{sig+}
\sigma_0(E) = {\Omega _{+}(u_0)}^{-1/2} .
\end{equation}
Therefore, in high-dimensional systems exponential localization
takes place only if the disorder is strong enough.

For $\sigma<\sigma_0(E)$ the function $H_{+}(z)$ has no poles, its
ROC its $|z| \geq 1$. This means that the unit circle $|z|=1$
belongs to the ROC, and the filter is stable.

We interpret solutions for this range of the disorder values,
$\sigma$, also as localized, however with non-exponential
localization. Non-exponential localization corresponds to a
Lyapunov exponent $\gamma=0$. This simply means that the
corresponding function grows slower than the exponent
$\exp(2\gamma n)$. To this class of functions do not only belong
power-law functions, but also exponents with a different argument,
e.g. $\exp(c\cdot n^{\delta})$ with $\delta <1$. Here we encounter
limits of applicability of the method, which come into play,
however, only in physically inaccessible systems of high
dimensionality. The elementary pole search can be applied only for
exponentially localized states, the general case requires a
detailed investigation of the filter. In its present form our
theory is not capable to determine the type of non-exponential
localization.

The curve for $\sigma_0(E)$ (Fig.1a) is the well-known mobility edge.
For $\sigma >\sigma_0(E)$ there exists an insulating phase,
whereas for $\sigma < \sigma_0(E)$ a metallic phase is found (bad metal).
A coexistence of phases is here not possible. Thus the phase transition
here has an appearance as if it were a transition of second order.
This simple idea, however,  is contradicted by the behaviour of the
Lyapunov exponent: the transition from $\gamma \equiv 0$ to $\gamma \neq 0$
is not continuous. One can clearly see that all these mobility edges
for high-dimensional systems show the same qualitative behaviour.

\begin{figure}[htbp]
  \begin{center}
    \epsfig{file=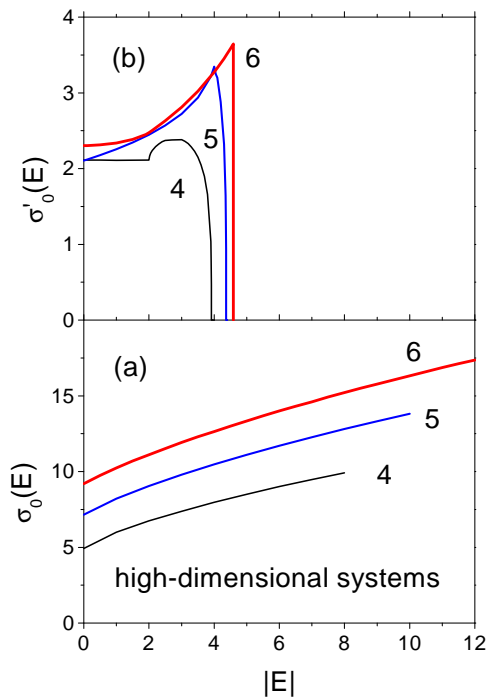,width=8cm}
    \caption{
     Threshold disorder values: (a) $\sigma_0(E)$  for the transition
     from the non-exponential to the exponential localization and (b)
     $\sigma^{\prime}_0(E)$ for the transition from
     the delocalized to the (non-exponentially) localized states.
     The curves are enumerated with the values of $D$.
      }
    \label{fig: 1}
  \end{center}
\end{figure}

\subsection{Stability and poles: solution $H_-(z)$}

Now let us turn to the second branch of the solution, $H_{-}(z)$.
In this case existence of the poles leads to principally different
consequences. Let us assume that the corresponding value of the
parameter $w$ is found and the pole $z=\lambda_1$ is located
inside the unit circle, $|\lambda_1|=1/\lambda$ with $\lambda >1$.
Formally, however, from the definition $w = -(z^{1/2}+z^{-1/2})$
the same value of $w$ can be obtained for $z=\lambda_2
=1/\lambda_1=\lambda$. The complex number $\lambda_2$ lies outside
the region of definition of the solution, $|z|\leq 1$. In this
sense the pole at $z=\lambda_2$ is virtual. For the inverse
$Z$-transform this fact is, however, irrelevant. The ROC for a
causal filter is defined by the inequality $|z|> \max|\lambda_i|$
or $|z|>\lambda>1$. Since the ROC and the region of definition of
the solution $|z|\leq 1$ do not intersect, a physical solution is
absent. Therefore, the filter $H_{-}(z)$ as a physical solution is
acceptable only if either there are no poles or they lie on the
unit circle. The latter case is realized for $D=2$ \cite{Kuzovkov}
and corresponds to so called marginal stability \cite{Weiss}. In
the following, we consider the general case $D\geq 4$ from a
unified point of view.

A pole of the first type is related to the singularity  of the
function eq.~(\ref{Hpm}) at $w=2$ (the root, for details see the
Appendix). We define the phase and amplitude via the integral
(\ref{Y})
\begin{eqnarray}\label{ph}
Y_D(w,E)=I_D(w,E)\exp(i\vartheta _D(w,E)) .
\end{eqnarray}
It is not difficult to show that this pole emerges at arbitrarily
small disorder for a negative phase $\vartheta _D(2,E)<0 $. The
equation $\vartheta _D(2,E_0)=0$ defines the boundary of the
region $|E|>E_0$, where the physical solution is absent and,
therefore, any disorder transforms the delocalized states into
localized ones. For high-dimensional systems the delocalized
states transform into states with non-exponential
localization. The corresponding $E_0$ values are $E_0=3.915$
(D=4), $E_0=4.365$ (D=5) and $E_0=4.578$ (D=6).

For the region $|E|<E_0$ there exists the physical solution with $\sigma
< \sigma^{\prime }_0(E)$, where $\sigma^{\prime }_0(E)$ is a second
threshold disorder value. The behaviour of this curve
$\sigma^{\prime }_0(E)$ (Fig.1b) is determined by resonance phenomena.
The integral in (\ref{Y}) consists asymptotically of a power function, $t^{(D-1)/2}$,
and a product of trigonometric functions. E.g. the function
$\cos(2t-\pi/4)$ comes from every Bessel function. If we represent this
product as a sum of monochromatic waves, then we denote the existence of a wave
with zero frequency as a resonance. For $w=0$ the first non-trivial resonance
lies either at $E_c=2$ (D=4,6,...) or at $E_c=4$ (D=5,7,...).

For $|E|<E_c$ a pole of the second type appearing at higher levels
of disorder $\sigma >\sigma_0^{\prime}(E)$ must be considered.
This type of pole emerges at purely imaginary values of the
parameter $w=iv$. It corresponds to the roots of the equation
\begin{equation}\label{omeg-}
  \sigma^2 \Omega _{-}(v)=1,
\end{equation}
where
\begin{equation}\label{W-}
  \Omega _{-}(v) = \frac{1}{\sqrt{ 4+v^2}} \int_0^\infty
\left[ J_0(2t)\right] ^{D-1} \cos(Et) \exp (-v t)dt.
\end{equation}
A physical solution is acceptable here only if the poles lie on the
unit circle, $|z|=1$ or $w \in [0,2]$ for our sector w. A marginally
stable solution corresponds to the value $w=0$ (or $v=0$ in eq.
(\ref{W-})). The threshold disorder value is given by
\begin{equation}\label{om-}
  \sigma^{\prime}_0(E) = {\Omega _{-}(0)}^{-1/2}.
\end{equation}
For $\sigma < \sigma^{\prime}_0(E)$ the function $H_{-}(z)$
does not possess poles, here we find the region of stability of the extended states.

Spaces of dimension $D=4,5$ possess a certain sensitivity with respect to the
resonance phenomena mentioned above. For
$E_c<|E|<E_0$ the line of the poles exhibits a deviation from purely
imaginary values of the parameter $w$ and touches the real axis at the
point $u^{\prime} \in [0,2]$. This point corresponds to the condition for the
phase of the integral $\vartheta_D(u^{\prime},E)=0$ and also belongs to the
unit circle (marginal stability). The corresponding threshold disorder value
we denote again as $\sigma^{\prime}_0(E)$:
\begin{equation}\label{sig}
\sigma^{\prime}_0(E)=\left
(\frac{\sqrt{4-{u^{\prime}}^2}}{I_D(u^{\prime},E)} \right )^{1/2}
.
\end{equation}
We see that the function $\sigma^{\prime}_0(E)$ is in general
singular in the energy. In going from the energy value
$E_c$ to the value $E_0$ the parameter $u^{\prime}$ increases monotonically
and reaches finally the value $u^{\prime}=2$. The function
$\sigma^{\prime}_0(E)$ goes continuously to zero for $|E|
\rightarrow E_0$.

For dimensions $D \geq 6$ there are no resonance phenomena. These
resonances are so weak that only one equation (\ref{om-}) remains
valid in the whole range of energies $|E|<E_0$. I. e. although
$D_c=4$ is a certain critical dimension (here non-exponential
localization arises), a qualitative agreement of all results is
only obrtained for $D \geq 6$.

In this way it emerges from our exact analytic theory that for
Anderson localization the upper critical dimension is $D_0=6$,
i.e. the problem in a certain way shows a similarity to
percolation \cite{Perc} - and this may not be totally unexpected.
However, we also note that as a rough estimate for the upper
critical dimension the value $D_c=4$ may also be accepted. It
corresponds to a first and important step of the qualitative
saturation of the results (the possibility of existence of
non-exponentially localized states), whereas a complete saturation
only obtains at D=6.

Quite generally in the range $|E|<E_0$ and under the condition
$\sigma < \sigma^{\prime}_0(E)$ there exists a region of existence
of stable delocalized states. Because in this range also non-localized states
coexist with the other ones, this means that in this range the
good metal (delocalized states) and the bad metal
(non-exponentially localized states) can form a heterogeneous system,
whose properties depend on the relative proportions of the subsystems;
e.g. the bad metal, if the subsystem of non-exponentially localized
states percolates. Fig. ~\ref{fig: 1} presents results of a
numerical solution of the resulting equations.


\section{Low-dimensional systems}

\subsection{Analytic solution}

Independently from whether $D_c=4$ or $D_0=6$ are taken as the
upper critical dimension, it is necessary to consider 3-D systems
as low dimensional ones. We approach here a widely held opinion
that all states in $D=3$ should be stable against perturbations
and should have a finite radius of convergence. This, however, is
not correct and would only be valid if the whole field of the
scaling theory of localization would be faultless.

The eq.(\ref{om+}) has always a solution for the physically
important cases $D=2,3$. We  refer to these cases as
'low-dimensional' ones. For the low-dimensional systems the
integral eq.~(\ref{W+}) can be evaluated analytically (the
formulas can be found in the tables of Laplace transforms). The
corresponding pole diagrams are shown graphically in the Appendix.
The system function has a pole at $z=\lambda=\exp(2\gamma)$ with
$\gamma > \gamma_0$, $2\cosh (\gamma_0) = u_0 = 2p + |E|$ and
$p=D-1$. Note, however, that the above mentioned feature of the
low-dimensional systems is caused by the divergence of the
integral $\Omega_{+}(u_0)$.

If one applies the obtained equations (\ref{om-}) for
$\sigma_0(E)$ to low-dimensional systems with $D=2,3$, then
formally $\sigma_0(E) \equiv 0$. Therefore, even infinitesimal
disorder leads to solutions with exponential localization. Because
the curve $\sigma_0(E)$ forms the border between exponentially
localized and non-exponentially localized states, this simply
means that in this model for low-dimensional systems
non-exponentially localized states are impossible (see, however, a
discussion below).
\begin{figure}[htbp]
  \begin{center}
    \epsfig{file=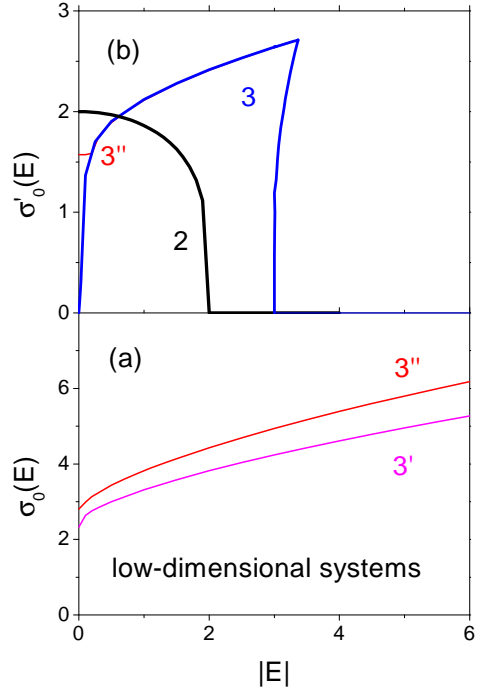,width=8cm}
    \caption{
     Threshold disorder values: (a) $\sigma_0(E)$  for the transition
     from the non-exponential to the exponential localization and (b)
     $\sigma^{\prime}_0(E)$ for the transition from
     the delocalized to the  localized states.
     The curves are enumerated with the values of $D$.
     The curves $3'$ and $3''$ correspond to the model
     eq.(\ref{Ekprime}) for $\kappa=0.01$ and $\kappa=0.1$.
      }
    \label{fig: 2}
  \end{center}
\end{figure}

For such systems the second curve  $\sigma^{\prime }_0(E)$ takes over the
role of the mobility edge. The shape of this curve (Fig.2.b) for low-dimensional
systems has a certain similarity with the ones for high-dimensional
systems. Here we must stress that $\sigma^{\prime }_0(E)$ does not represent
a true mobility edge, but an upper limit for the coexistence of the two phases.

For $D=2$ in \cite{Kuzovkov} it has been found analytically that
$\sigma^{\prime }_{0}(E)= 2 (1-E^2/4)^{1/4} $ for $E < E_0$ where
$E_0=2=D$. I.e. in this case the value $E_0$ corresponds exactly to the midst
of the band half-width $2D=4$. For $E > E_0$ any disorder transforms the
delocalized states into localized ones.

For $D=3$ the value $E_0=3=D$ corresponds again to the midst of the
band half-width $2D=6$. The resonance value mentioned above $E_c$ lies
for $D=3$ at $E_c=4$. Because $E_0<E_c$, the resonance does not play a
role here. The mobility edge $\sigma^{\prime }_{0}(E)$ again follows from
eq. (\ref{om-}).

The case of the 3-D system shows an exception in the energy range
$E_0<|E|<E^{\prime}_0=3.367$, where the function $\sigma^{\prime
}_0(E)$ exhibits so-called reentrant behaviour. The proof of this
requires a special investigation (see pole diagrams in the
Appendix).

Note that for the 3-D system in the band center $\sigma_{0}(0)=0$,
because the integral in eq.~(\ref{W-}) diverges. Fig. ~\ref{fig:
2} presents results of a numerical solution of the resulting
equations. These results are compared to the analytical result of
\cite{Kuzovkov} for the two-dimensional system (curve 2).

\subsection{Phase diagram and logarithmic divergence}

Although the results for $D=3$ are formally exact, they require a special
discussion. The mathematically correct results are physically
acceptable only if they are stable against small perturbations
(e.g. small changes in the model definition, fluctuations of parameters
in the model). From this point of view two results have to be questioned:
(i) the singularity of $\sigma^{\prime }_{0}(E)$ in the band center,
$\sigma_{0}(0)=0$, and (ii) the non-existence of non-localized
states, $\sigma_0(E) \equiv 0$. Point (i) arises mathematically
as a consequence of the logarithmic divergence of the
integral (\ref{W-}) for $v=0$ and $|E|\rightarrow 0$. Point
(ii) arises again via a logarithmic divergence of the
integral (\ref{W+}) for $u=u_0$. I.e. basically, the case
$D=3$ is nothing else but a type of logarithmic deviation from the
high-dimensional case. Under certain conditions one can find the perturbations
which are capable of regularizing the mentioned
logarithmic divergence, i.e. to transform them into a finite term.
One could e.g. surmise that in this regularization
correlated disorder \cite{Moura98} might play an important role.
One must stress here that we are dealing only with results for
tight-binding Hamiltonians with diagonal disorder. It is largely unclear,
whether this property remains valid also for non-diagonal disorder.

To illustrate this let us first consider a purely mathematical
problem. Is it at all possible to confirm the results (i) and (ii)
either by different numerical computations or analytical
approaches? The answer is - no! Any deviation from the exact
solution (via numerical or analytical ways) automatically
generates results of a mean-field theory. A mean-field theory has
a certain qualitative agreement with the results of the exact
theory, but only for high-dimensional systems. Everything looks as
if any uncontrolled deviation from the exact theory has added
additional dimensions to a 3-D system. As a mathematical model let
us further consider a 4-D system, where, however, a coupling
involving this additional dimension is exceedingly weak (parameter
$\kappa\ll 1$). We start from a generalization of the function
(\ref{Ek}) for $D=3$:
\begin{eqnarray}\label{Ekprime}
\mathcal{E}(\mathbf{k})=E - 2\sum_{j=1}^2\cos (k_j)-2\kappa \cos
(k_3).
\end{eqnarray}
Here the term involving $\kappa$ corresponds to the hopping matrix
element into the fourth dimension. Fig.\ref{fig: 2} gives the
numerical results for 2 cases: $\kappa=0.01$ (curve $3'$) and
$\kappa=0.1$ (curve $3''$). One clearly sees that all results,
which have nothing to do with the logarithmic divergence, e.g. the
entire curve $\sigma^{\prime }_{0}(E)$ with the exception of the
point $E=0$, remain extremely stable. Even the change in the
parameter $E_0$ is of the order of $O(\kappa^2)$.

The 'logarithmic' results on the other hand turn out to be
completely unstable. Even the small values of the coupling (or
regularization) parameter $\kappa$ produce results which are in
qualitative agreement with those for high-dimensional systems. The
curves $3'$ and $3''$ look as if they were an extrapolation of the
corresponding curves for $\sigma_{0}(E)$ from Fig.~\ref{fig: 1}a.
Here the inequality $\sigma_{0}(E)> \sigma^{\prime }_{0}(E) $ also
applies. Reducing the parameter $\kappa$ slowly moves the mobility
edge $\sigma_{0}(E)$ downwards, because the dependence on the
parameter $\kappa$ is extremely week (logarithmic), $\sigma_{0}(E)
\sim 1/\ln(\kappa^{-1})$. Only for even smaller values of the
parameter $\kappa \ll 0.01$ one might perhaps see traces of the
exact results, because in this case one has $\sigma_{0}(E)<
\sigma^{\prime }_{0}(E) $ for $|E|<E_0$.

\section{Conclusion}

Although the formal investigation of the Anderson model of
localization (tight-binding Hamiltonian with diagonal disorder)
for higher spatial dimensions D might at first look very abstract,
the exact analytical results supply us with clear physical
consequences. The Anderson problem of localization and the
percolation problem belong to the same class of critical
phenomena: both have the same lower and upper critical dimensions.
I.e., although the Anderson model appears to be much more complex
and richer, certain fundamental results appear to be transferable.
Percolation is possible for 2-D systems, this corresponds to the
existence of a metal-insulator transition in disordered 2-D
systems. In this sense there is no reason to believe that the
existing contradictions between theory and experiment for 2-D
systems point to an incompleteness of the Anderson model. On the
contrary, our analytical investigation has shown that a
tight-binding Hamiltonian is presumably sufficient for this
purpose.

The main problem of the theory thus does not rest in the
Hamiltonian, but rather in the interpretation of the results,
which mainly derive from numerical work. Our analytical and exact
results demonstrate the necessity of interpreting the phase
transition in the framework of first order phase transition theory
and this holds independently of the spatial dimension $D\geq 2$.
If one, however, attempts to apply procedures which have only been
developed for systems with a second order phase transition (and
this is the general case), one does not necessarily obtain wrong
numbers, but an incomplete or even wrong interpretation. See an
example in \cite{Kuzovkov}, where an analytical (exact) scaling
function for the 2-D system has the same form as obtained by
numerical scaling. The physical interpretation is, however totally
different.  We hope that the necessary corrections of the
numerical tools are possible to detect the first order phase
transition.

The rather large value of the upper critical dimension for the Anderson
localization (and the percolation) problem permits to consider 2-D
and 3-D cases as low-dimensional systems. Thus a revisiting
of the results is also necessary for 3-D systems.
We have found that the 3-D case is nothing else
but a type of logarithmic deviation from the high-dimensional
case. As a consequence results a certain instability of the results,
whose details are discussed in the text.

We give here a short summary of the main results.

\begin{itemize}
\item For the Anderson localization problem there exists an upper
critical dimension $D_0=6$. This value is also characteristic of a
related problem: percolation \cite{Perc}. For $D\geq D_0$ all
phase diagrams are qualitatively the same, only the corresponding
critical values develop in a monotonic way. One can also say that
this is the property of a mean-field theory, although in this case
a mean-field theory does not exist as a closed theory.

\item There exists also a second upper critical dimension $D_c=4$,
which has a different meaning. The states with non-exponential
localization are formed only for $D\geq D_c$, whereas for $D<D_c$
localized states are always exponentially localized. This second
upper critical dimension $D_c$ divides the dimensions into two
classes: high dimensions with $D\geq D_c$ and low dimensions with
$D < D_c$.

\end{itemize}

The results for the nontrivial spatial dimensions $D>1$ can be
summarized as follows.

(i) All states with energies $|E| > E_0$ are localized at
arbitrarily weak disorder. The value of $E_0$ depends on $D$ and
lies inside the band $E_0 < 2D$.

(ii) For $|E|<E_0$ states are only localized if the disorder
$\sigma$ exceeds a critical value $\sigma^{\prime}_0(E)$,
otherwise a two-phase system is formed from an insulating and a
metallic one. This differs from the traditional point of view
which considers the localization transition as a continuous
(second order) transition. Should the standard interpretation of
this system in the framework of first-order phase transition
theory be applicable (which still has to be investigated) one can
expect that qualitatively it has similar properties as other
heterogeneous two-phase systems (e.g. the coexistence of water and
ice). Then percolation problems might be important.

(iii) Exponential localization always exists for the physically
important cases $D=2,3$. Non-exponential localization occurs only
for higher dimensions  $D \geq D_c=4$. In this case for $|E|<E_0$
and $\sigma<\sigma^{\prime}_0(E)$ first the heterogeneous system
appears, where the difference between the two phases may be small
(this has to be investigated). For
$\sigma_0(E)>\sigma>\sigma^{\prime}_0(E)$ one finds a homogeneous
system with non-exponential localization. Only for
$\sigma>\sigma_0(E)$ a system with exponential localization
appears.

(iv) $\sigma^{\prime}_0(E)$ is in general not an analytic function
of the energy $E$. There exist certain resonances.

\begin{acknowledgments}
V.N.K. gratefully acknowledges the support of the Deutsche
Forschungsgemeinschaft.
\end{acknowledgments}

\appendix
\section{Pole diagrams}
\subsection*{Parametric representation of the pole diagram}

A filter $h_n$ is characterized by a pole diagram of its image
$H(z)$. The principal definition eq.(\ref{H}), together with the two
other ones, eqs. (\ref{rechts2}),(\ref{zz}), supply us with the parametric
w-representation of the pole diagram, eq.(\ref{Hpm}).
Let us rewrite this relation into the form
\begin{equation}\label{a100}
    \frac{1}{H(z)}=1-\sigma^2 R(w,E) ,
\end{equation}
where $R(w,E)$ is generally a complex function of the complex
variable $w$ and energy $E$. The main idea is quite simple.
The function $H(z)$ has its poles where
\begin{equation}\label{a101}
\sigma^2 R(w,E) \equiv 1 .
\end{equation}
An elementary requirement for this is the condition on the
argument of the complex function
\begin{equation}\label{a102}
\arg R(w,E) = 0 ,
\end{equation}
because $\sigma$ is positive. For the given energy value $E$
eq.(\ref{a102}) defines one or more lines (pole lines) in the
complex $w$-plain. As we have already discussed in the text, for
reasons of symmetry it suffices to analyse only a sector $w=u+iv$
with $u \geq 0$ and $v \geq 0$. Let us analyse one of these pole
lines. For each point $w$ on this line the position of the pole is
determined via eq.(\ref{z-inverse}). The corresponding value of
the disorder $\sigma$ is found from eq.(\ref{a101}):
\begin{equation}\label{a103}
\sigma=R(w,E)^{-1/2} .
\end{equation}
Because for each value of $w$ on the pole line there is associated a
value of $\sigma$, it is possible to indicate by an arrow next to
the line in the diagram  in which direction the poles move with
increasing disorder.

\subsection*{Filter $H_{+}(z)$}

Let us consider first the simplest case, the filter
$H(z)=H_{+}(z)$. Here the poles are connected with the notion
of the Lyapunov exponent. In the parametric representation
there exists a simple relation
\begin{equation}\label{a105}
w=2 \cosh(\gamma) ;
\end{equation}
thus the eqs.(\ref{a103}),(\ref{a105}) together supply a connection
between $\sigma$ and $\gamma$.

For a  2-D system it is possible to analytically evaluate the
corresponding function
$R(w,E)$  \cite{Kuzovkov}:
\begin{equation}\label{a104}
R(w,E)=\frac{1}{2\sqrt{w^2-4}}[\frac{1}{\sqrt{(w+E)^2-4}}
+\frac{1}{\sqrt{(w-E)^2-4}}] ,
\end{equation}
or
\begin{eqnarray}\label{a106}
R(w,E)=\frac{1}{2\sqrt{w^2-4}} \times \\ \nonumber
[\frac{1}{\sqrt{(w+E+2)(w+E-2)}} +\frac{1}{\sqrt{(w-E+2)(w-E-2)}}]
.
\end{eqnarray}

We clearly see that the function (\ref{a106}) possesses for the given energy
several points, where it diverges. These values $w=-2-E$, $w=2-E$, $w=E-2$
and $w=E+2$ are real. They are nothing else but the resonances discussed
in the text. To detect these we investigate the $w$-parametric
representation of the integral eq.(\ref{Y}).
Every Bessel function $J_0(2t)$ contributes asymptotically a
trigonometric function, $\cos(2t-\pi/4)$. In addition there exists
another energy dependent trigonometric function,
$\cos(Et)$. If we represent all these functions via complex exponentials,
we obtain under the integral in
eq.(\ref{Y}) asymptotically a product of the power function and
a sum (with well determined coefficients) of exponents
$\exp[i(w-w_j)t]$. These are the resonance values $w_j$. In the 2-D
case we have a strong resonance (the function $R(w_j,E)$
diverges).

It is easy to establish that the function $R(w,E)$ satisfies
eq.(\ref{a102}) for the real value of the parameter $w$ under the condition
\begin{equation}\label{a107}
w \geq u_0=\max \{w_j\} ,
\end{equation}
where for the 2-D system $u_0=2+|E|$. I.e. here exists a line of poles
which starts from the point $u_0=2+|E|$ (it corresponds to the value
$\sigma=0$, because here $R(u_0,E)^{-1/2}=0$) and moves on with
increasing value of the parameter $\sigma$ always along the real axis.
It is also easy to find that in this case no other pole lines exist.
This result can be generalized for higher dimensions. So one can establish
that the value of $u_0$ found from eq.(\ref{u0}) corresponds
precisely to the condition eq.(\ref{a107}): we always have to deal with the same resonance.

\begin{figure}[htbp]
  \begin{center}
    \fbox{\epsfig{file=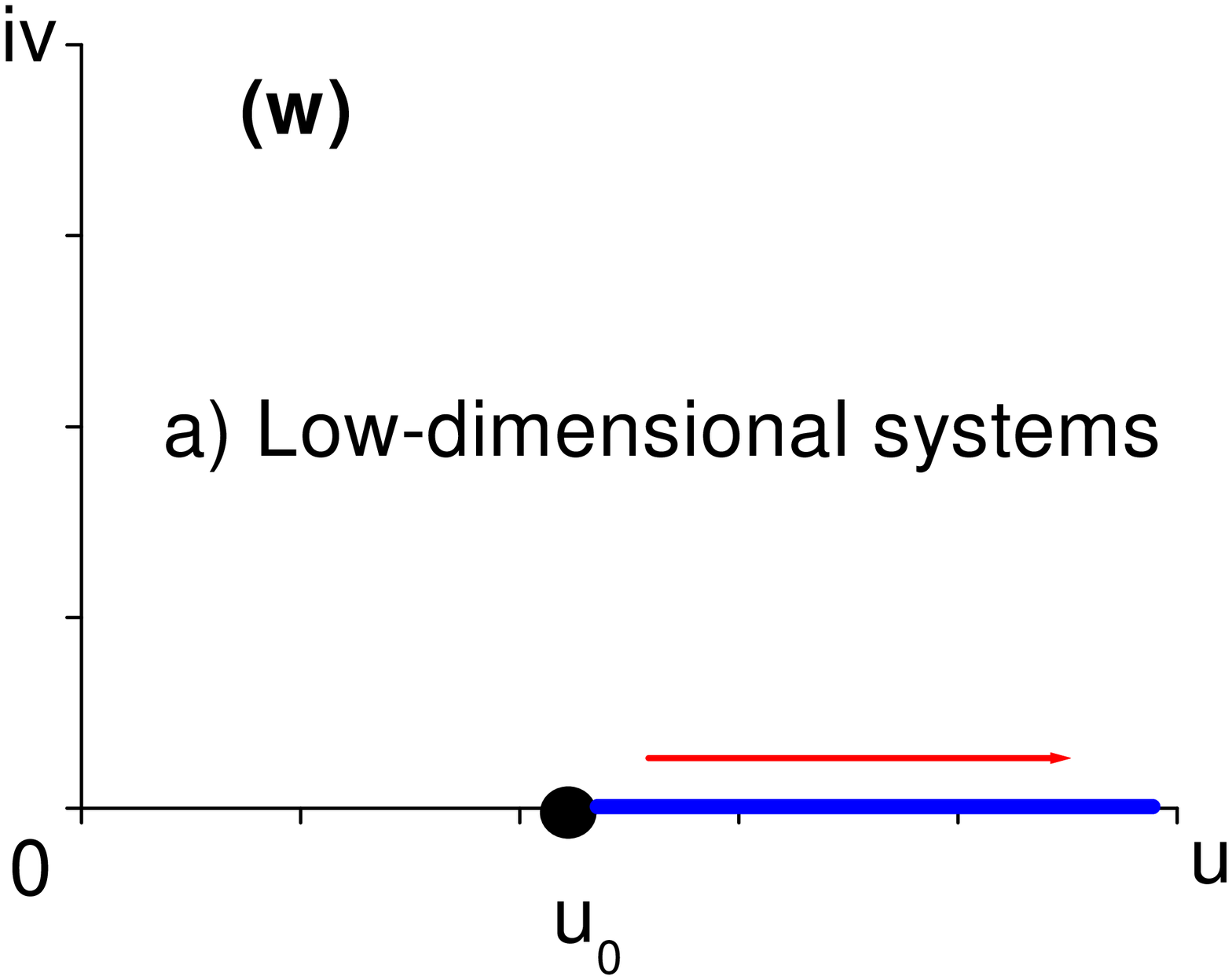,angle=0,width=5cm}}
    \fbox{\epsfig{file=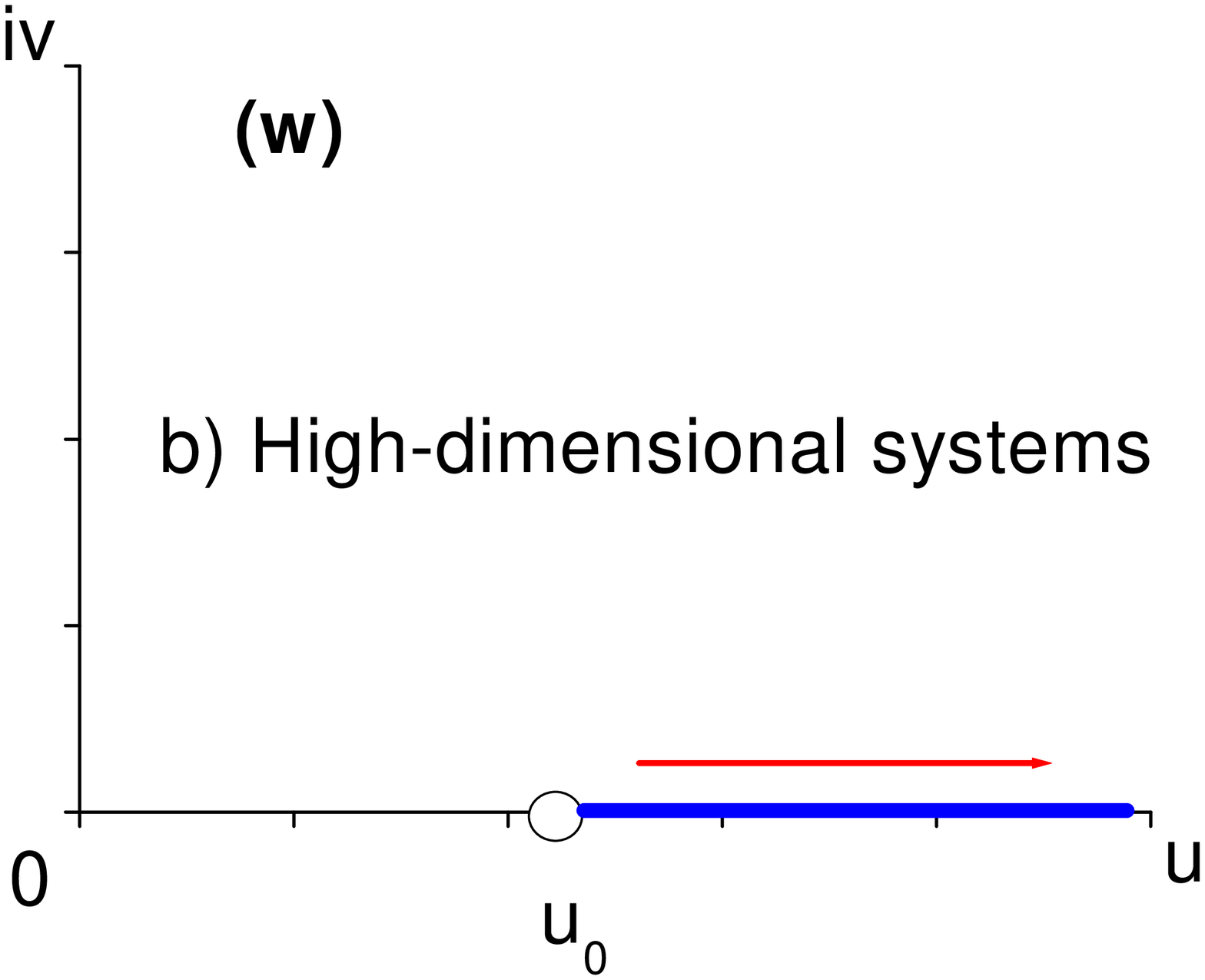,angle=0,width=5cm}}
    \caption{Parametric representation of the pole diagram for an
    unstable filter: a) low-dimensional systems with $D=2,3$; b)
    high-dimensional systems with $D \geq 4$.
      }
    \label{fig:3}
  \end{center}
\end{figure}

The structure of the pole diagram is nearly identical for all
dimensions $D$, Fig.\ref{fig:3}. There exists only one pole line,
which originates at the point $w=u_0$. The direction of increasing
$\sigma$ is denoted by an arrow. The only difference is that for
low-dimensional systems the starting point $u_0$ corresponds exactly to
$\sigma=0$ (this is marked in the diagram with a black circle),
for high-dimensional systems the value
$\sigma_0(E)=R(u_0,E)^{-1/2}$ is finite (the point $w=u_0$ is
marked with a white circle). In the latter case it is impossible to find
for smaller values of the disorder $\sigma<\sigma_0(E)$ a point
in the diagram, where the condition eq.(\ref{a101}) is fulfilled. The
filter $H_{+}(z)$ possesses no poles. The border thus found via eq.(\ref{sig+})
has been defined as mobility edge.

\subsection*{Filter $H_{-}(z)$}

This filter has a physical interpretation only under the condition
\cite{Kuzovkov} that the function $H_{-}(z)$ has either no poles
or that they belong to the unit circle, $|z|=1$. In the
parametric w-representation the unit circle (in the sector
$w=u+iv$ with $u\geq 0$, $v\geq 0$) corresponds to an interval
on the real axis $u\in[0,2]$. I.e. it is necessary first to find the
single points or even lines in this interval $u\in[0,2]$,
where the condition eq.(\ref{a102}) holds. Later on one must also find out,
in which way the pole line leaves this interval.

It is easy to ascertain that one of the possible pole lines always
lies on the imaginary axis $v$, and in the interval $v\geq 0$. On
this line (trivial pole line) only the point $v=0$ could have a
physical interpretation, because the point, $w=0$, also belongs
to the unit circle. The existence of other pole lines strongly
depends on the space dimension $D$ and the energy $E$.

\begin{figure}[htbp]
  \begin{center}
    \fbox{\epsfig{file=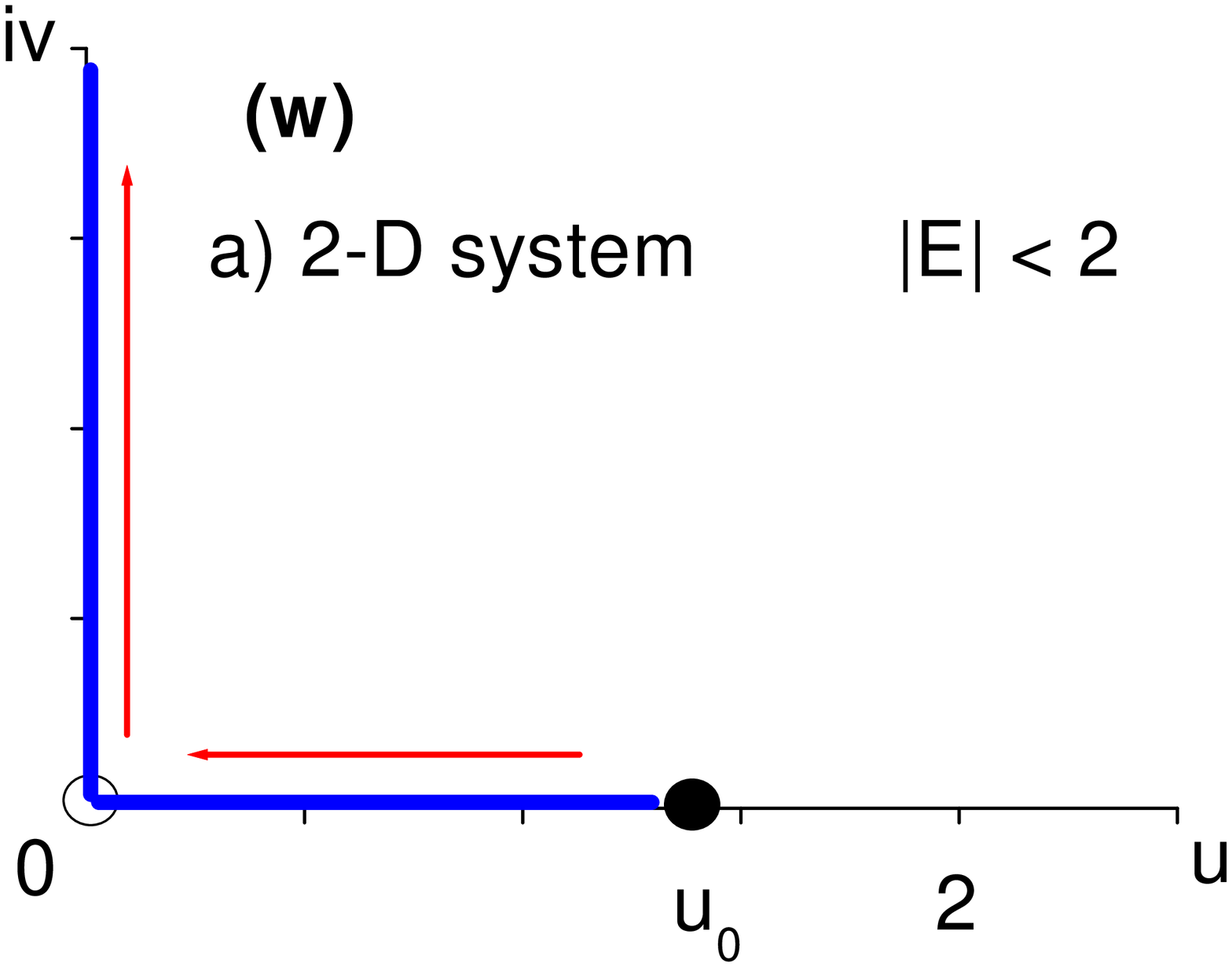,angle=0,width=5cm}}
    \fbox{\epsfig{file=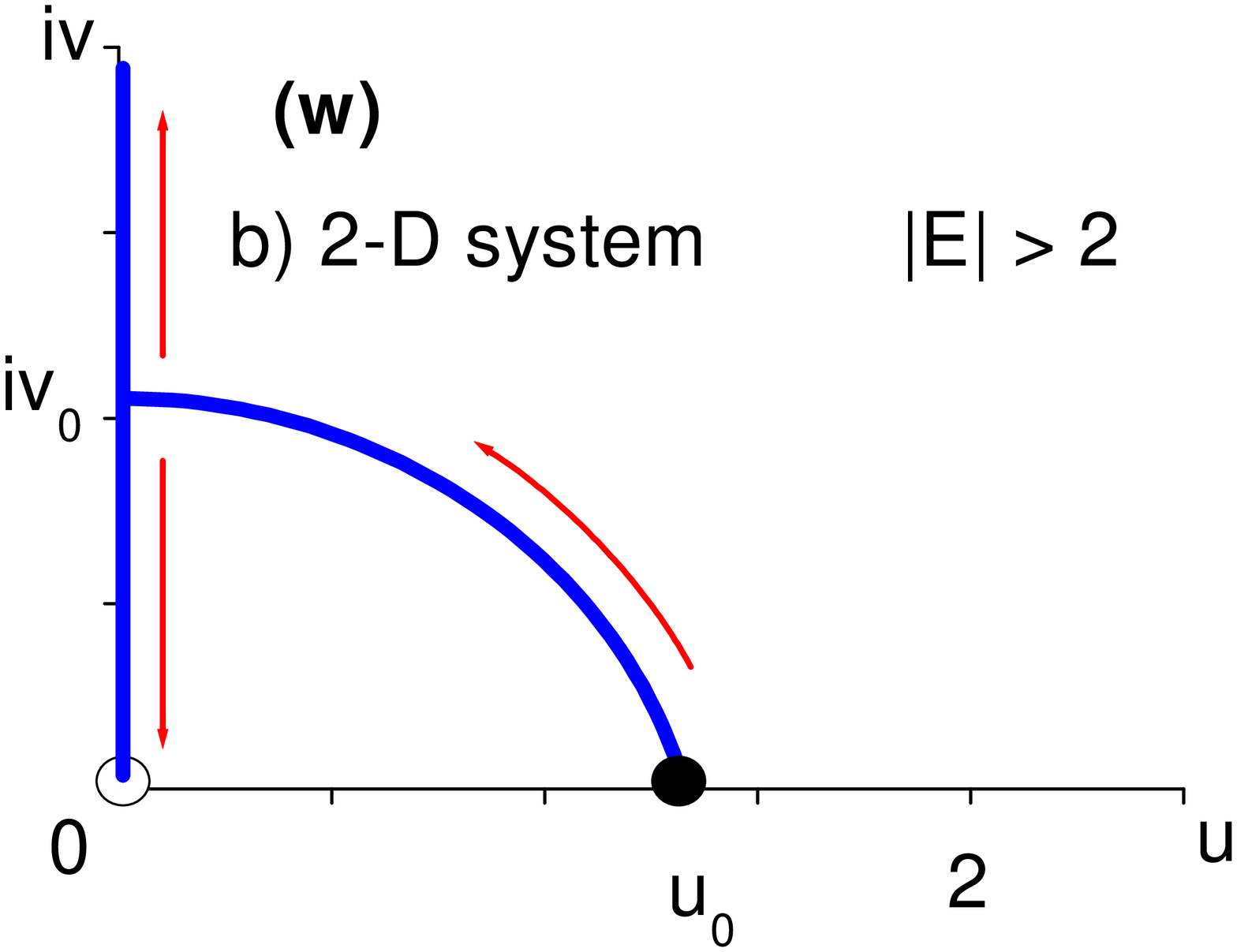,angle=0,width=5cm}}
    \caption{Parametric representation of the pole diagram for a 2-D
    system: a) energy range $0\leq |E| < 2$; b) energy range $2
    <|E|<4$.
      }
    \label{fig:4}
  \end{center}
\end{figure}

As the simplest example let us first consider the filter for the 2-D
system \cite{Kuzovkov}, where the corresponding function has an analytical form:
\begin{equation}\label{a108}
R(w,E)=\frac{1}{2\sqrt{4-w^2}}[\frac{1}{\sqrt{4-(w+E)^2}}
+\frac{1}{\sqrt{4-(w-E)^2}}] .
\end{equation}
Because the interval $u\in[0,2]$ plays an important role for the physical
interpretation, we consider first the resonance values
$\{w_j\}$, which lie precisely in this region. It is easy to calculate
that for each energy value only one resonance $w=u_0$
exists. For energies  $0\leq |E| < 2$ we have $u_0=2-|E|$.
In the remaining energy region, $2 <|E|<4$, one has
$u_0=|E|-2$. In the 2-D system the resonances are so strong (the
function $R(w,E)$ diverges at $w=u_0$) that these determine
in a unique manner the beginning of the new pole lines and their direction,
Fig.\ref{fig:4}.

In the interval $0\leq |E| < 2$ the pole line starts exactly
at the point $u_0=2-|E|$. This point corresponds to the value of the
disorder $\sigma=0$ and is denoted by a black circle in the diagram.
With increasing value of $\sigma$ the pole line follows exactly the real axis
$u$, until it reaches the point $u=0$ ($w=0$). At this point the
corresponding disorder $ R(0,E)^{-1/2}$ is finite (white circle in the figure),
and we have denoted this in eq.(\ref{om-}) as
$\sigma^{\prime}_0(E)$. Because this line is always a unit circle
(marginal stability), we have physical solutions.
These exist only under the condition that $0\leq \sigma < \sigma^{\prime}_0(E)$.
If the disorder crosses the border $\sigma^{\prime}_0(E)$,
the pole line leaves the point $w=0$ and follows further the trivial
pole line (imaginary axis). Here, however, a physical interpretation is
no longer possible and consequently there are no extended states in the range
$\sigma>\sigma^{\prime}_0(E)$.

In the interval $2 <|E|<4$ the behaviour of the pole line determines
the resonance at $w=u_0=|E|-2$. This line again starts at the value
$u_0$, which corresponds to the disorder $\sigma=0$ (black circle
in the figure). Then this line leaves the real axis and takes its course
into the complex plain until it reaches a particular point $v_0$
(bifurcation point) on the imaginary axis $v$.
The later path makes use of branches of trivial pole lines.
The directions belonging thereto are again marked with an arrow.
Thus we have for $\sigma >0$ no points in the phase diagram
which can be interpreted physically. I.e.
an infinitesimal disorder already suffices in this energy range
to destroy all extended states.
One may also write that in this range $\sigma^{\prime}_0(E)\equiv 0$.

Formally between the 3-D case (low-dimensional system) and the
cases $D \geq 4$ (high-dimensional systems) there is a
quantitative difference. For high-dimensional systems one does not
find a divergence of the function $R(w,E)$ at the resonance
$w=u_0$, but only a jump in the argument of this complex function.
For the 3-D system this divergence exists, on the other hand, but
in contrast to the 2-D system the resonance is very weak
(logarithmic divergence). In all these cases the resonances have
no direct influence: they do not generate a pole line emerging
from the point $u_0$. There is, however, an indirect influence of
the resonances, because every resonance defines the argument of
the function $R(w,E)$.

\begin{figure}[htbp]
  \begin{center}
    \fbox{\epsfig{file=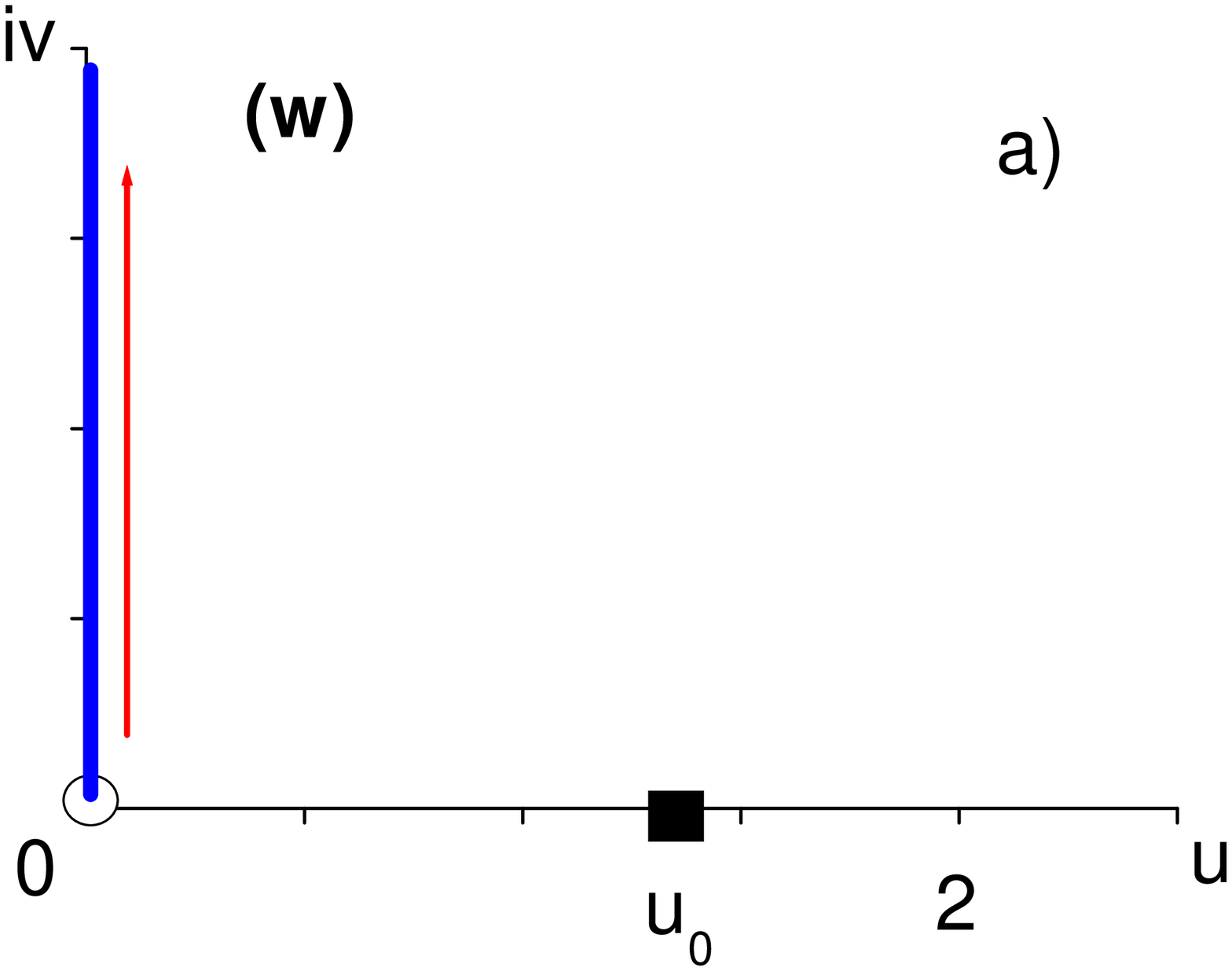,angle=0,width=5cm}}
    \fbox{\epsfig{file=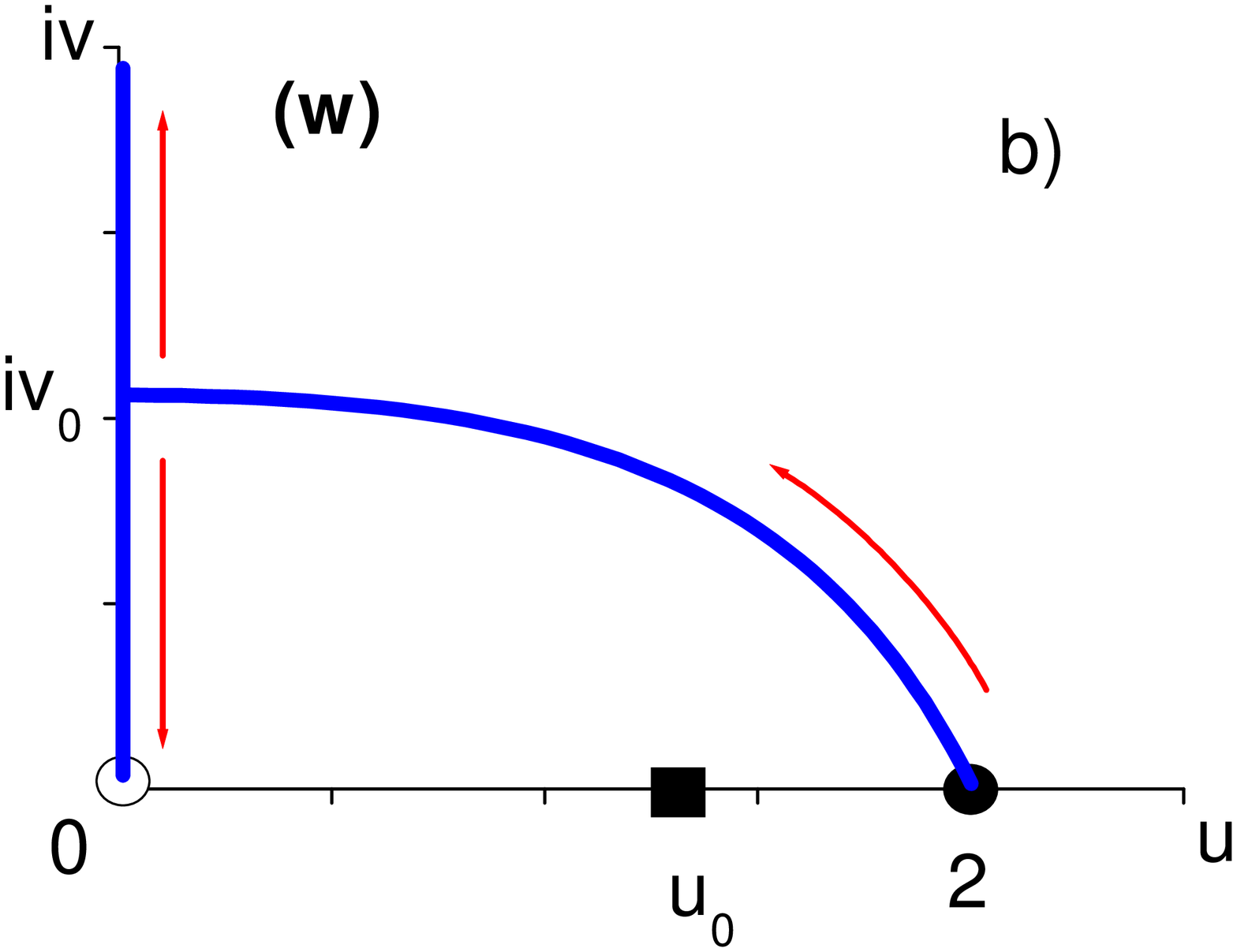,angle=0,width=5cm}} \\
    \fbox{\epsfig{file=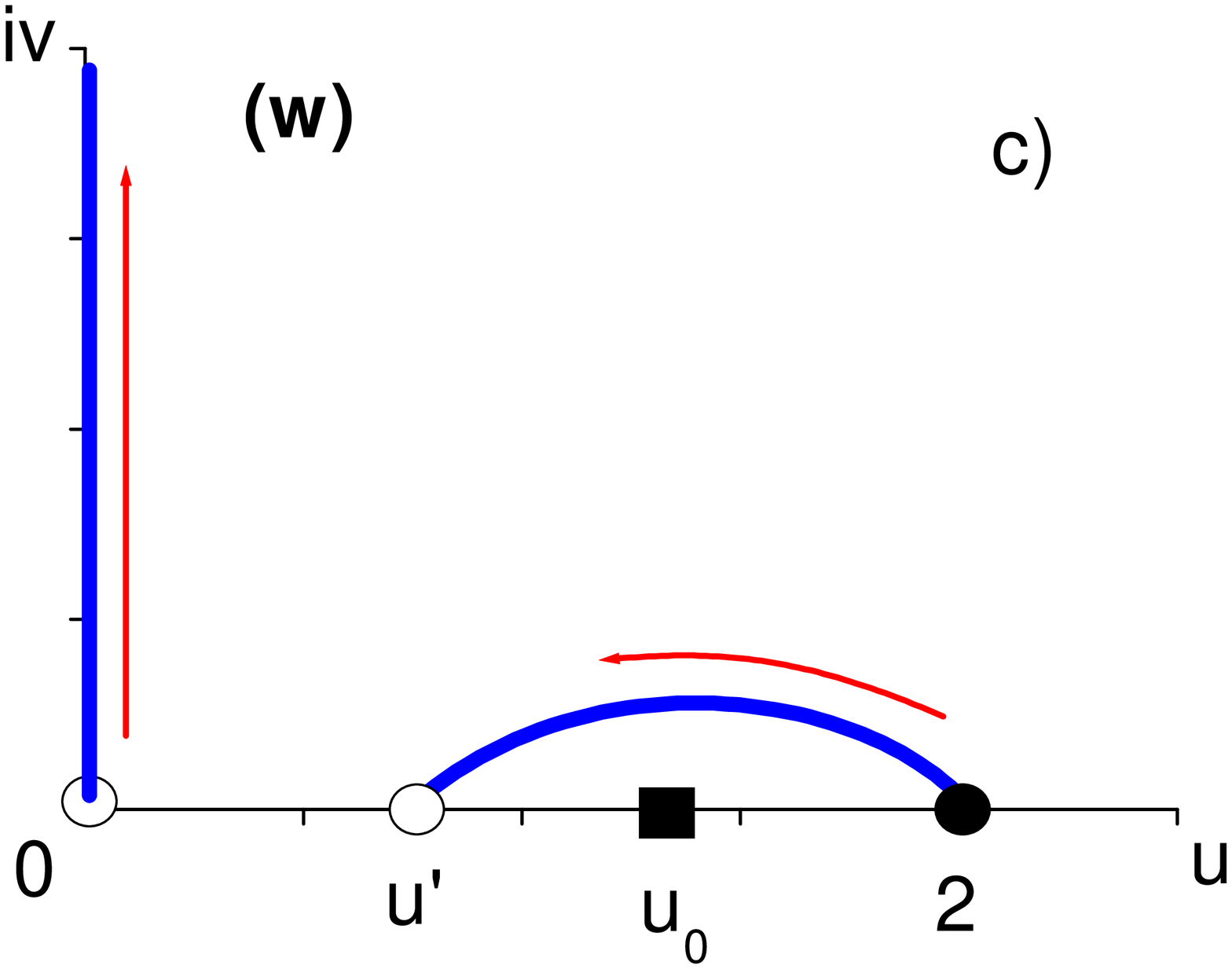,angle=0,width=5cm}}
    \fbox{\epsfig{file=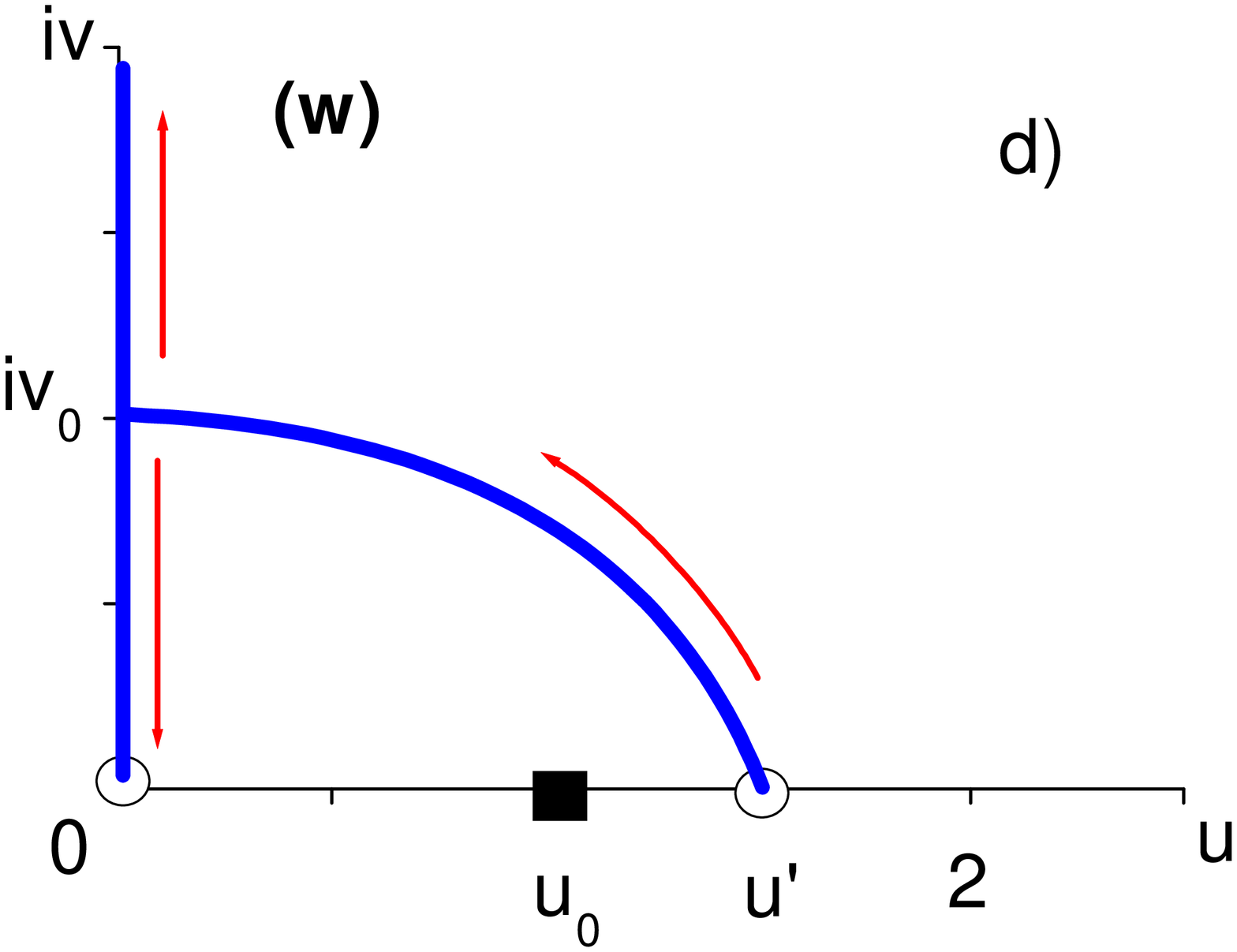,angle=0,width=5cm}}
    \caption{Different types of pole diagrams for the systems
    with spatial dimensions $D \geq 3$. For details see the text.
      }
    \label{fig:5}
  \end{center}
\end{figure}

In fig.\ref{fig:5}, different types of pole diagrams are
presented. Fig.\ref{fig:5}a is typical for small values of the
energy. Although the resonance at $w=u_0$ exists (this point is
marked in the figure with a black square), the weak resonance does
not generate a new pole line. There remains only the trivial pole
line, where only the point $w=0$ has physical significance. The
value $ R(0,E)^{-1/2}$ defines again the function
$\sigma^{\prime}_0(E)$ already mentioned. Physical solutions exist
only for $\sigma < \sigma^{\prime}_0(E)$ (no poles on the unit
circle). Fig.\ref{fig:5}b is typical for large values of the
energy. In the parametric representation of the filter function,
eq.(\ref{Hpm}), one can in addition see a point, where the
function $R(w,E)$ diverges. This is the point $w=2$ (a type of
energy independent resonance, which corresponds to the factor
$\sqrt{4-w^2}$). One can see that this point takes over the role of
the resonance at $u_0$. The pole line starts from the point $w=2$
(this point corresponds to the value $\sigma=0$ and is therefore
again marked with a black circle). The continuation is similar to
Fig.\ref{fig:4}b: the line continues to the imaginary axis and
reaches it at the point $v=v_0$. Although the points $w=u_0$ and
$w=2$ are not identical, Fig.\ref{fig:4}b and Fig.\ref{fig:5}b are
qualitatively similar, and they have the same physical
interpretation. In this range of the energy an infinitesimal
disorder destroys all extended states.

An indirect influence by the resonances consists in the fact that one
can change in their neighbourhood the argument of the complex function $R(w,E)$
in such a way that the condition eq.(\ref{a102}) becomes valid at a point
$w=u'$, see Fig.\ref{fig:5}c  and Fig.\ref{fig:5}d.
The point $u'$ is not a resonance, here the function
$R(u',E)$ remains finite (white circle in the pole diagram).

Let us consider further on two examples. In the 3-D case one can
divide the energy values into three ranges. In the range $0\leq
|E|<2$ the resonance is found at $u_0=|E|$, which, however, has no
importance. We have the pole diagram of Fig.\ref{fig:5}a. In the
range $2\leq |E|<4$ the resonance occurs at $u_0=4-|E|$. Here the
above mentioned point $u'$ arises, but under the condition $|E| >
E_0=3.00$. I.e. in the range $2\leq |E|<E_0$ the pole diagram of
Fig.\ref{fig:5}a remains valid. In the range $E_0 < |E|<4$ we find
a different typ of diagram, Fig.\ref{fig:5}c. The point $u'$
corresponds to the condition $0\leq u'<u_0$. For $E\rightarrow 4$
the point $u'$ moves towards the value $u'=0$, and thus arises the
bridge between the new pole line and the trivial pole line. After
this the diagram is qualitatively the same as Fig.\ref{fig:5}b:
there are no extended states. This is also valid in the range $4 <
|E|<6$, where $u_0=|E|-4$.

Figure \ref{fig:5}c describes a complicated case which does not have an
unambiguous interpretation. Formally this is the only diagram which has
two pole lines. The other ones consisted always of a single pole line,
although bifurcation points were also possible. For infinitesimally small
disorder the resonance at $w=2$ is important, as
in the diagram of Fig.\ref{fig:5}b. Here with increasing disorder $\sigma$,
the pole line which starts at the point $w=2$ (black circle), leaves the
interval $u\in [0,2]$. The corresponding poles have no physical interpretation,
which corresponds to the annihilation of extended states via infinitesimal disorder.
If the disorder increases further this pole line approaches again the
interval $u\in [0,2]$, and reaches it at the point $u'$ (Fig. \ref{fig:5}c).
The corresponding value of the disorder
$\sigma_1=R(u',E)^{-1/2}$ is finite (white circle).
The physical interpretation now depends to which value of the disorder
$\sigma_2=R(0,E)^{-1/2}$ corresponds the point $w=0$ (white
circle) on the trivial pole line.

If $\sigma_1 < \sigma_2$, then a sort of gap arises in the
disorder in such a manner that in the range
$\sigma_1<\sigma<\sigma_2$ no values of the parameter $w$
correspond to the pole line. I.e. the filter $H(z)$ has no poles
in this range. Consequently a physical interpretation of the
solution is possible here and we obtain the reappearance of
extended states at finite disorder values in this special
interval. This condition, $\sigma_1 < \sigma_2$, is valid for 3-D
systems only in a very narrow range of the energy,
$E_0<|E|<E^{\prime}_0$, where $E^{\prime}_0=3.367$. The behaviour
of the curve $\sigma^{\prime}_0(E)$ is shown in Fig.\ref{fig: 2}
(so-called reentrant behaviour). We can clearly see that to each
energy value in the range $E_0<|E|<E^{\prime}_0$ are associated
three values of $\sigma^{\prime}_0(E)$; these are in particular
$\sigma^{\prime}_0(E)=0$ - localization via infinitesimal
disorder, $\sigma^{\prime}_0(E)=\sigma_1$ - reappearance of
extended states, and $\sigma^{\prime}_0(E)=\sigma_2$ - again
localization.

Outside this range, $E^{\prime}_0<|E|<4$, one has $\sigma_1 >
\sigma_2$. For this condition there are no points on the pole line which
permit a physical interpretation. I.e. after this annihilation of extended states
via infinitesimal disorder it is not possible for extended states to
reappear.

In the 4-D case we find a different sequence of resonances. In the
range $0 \leq |E|<2$ the resonance occurs at $u_0=2-|E|$, and the
diagram Fig.\ref{fig:5}a is valid. In the range $2 < |E|<4$ the
resonance is found at $u_0=|E|-2$. Here the point $u'$ ($u_0 > u'
\geq 2$) arises. The pole diagram corresponds to \ref{fig:5}d. The
pole line starts from a white circle, which corresponds to the
value $\sigma^{\prime}_0(E)>0$. Only this point in the pole
diagram has a physical interpretation, and it means the existence of
extended states under the condition of small values of the
disorder parameter, $\sigma<\sigma^{\prime}_0(E)$. For
$E\rightarrow 4$ the values $u'$ and $u_0$ move towards $w=2$.
Because always $u'>u_0$, this means that the point $u'$ reaches
the value $w=2$ before the point $u_0$, and this occurs for
$E_0=3.915 <4$. For $E > E_0$  the diagram \ref{fig:5}b is
still valid, although the series of resonances is not yet
exhausted: a resonance arises at $u_0=6-|E|$ for $4 < |E| <6$, and another one
at $u_0=|E|-6$ for $6<|E|<8$. Here also extended states are
not possible.


\end{document}